УДК 004.891.3, 004.93'14

# ПРИМЕНЕНИЕ НЕЙРОННЫХ СЕТЕЙ ДЛЯ ВЫЯВЛЕНИЯ ТРОЛЛЕЙ В СОЦИАЛЬНЫХ СЕТЯХ

А.В. Филимонов, А.В. Осипов, А.Б. Климов

## Введение

Влияние средств массовой информации на формирование национальных, политических и религиозных взглядов населения неоспоримо. Ограниченность объемов, авторская ответственность позволяет жестко контролировать весь информационный материал, который печатается в газетах, озвучивается на радио и телевидении.

В последние годы социальные сети получили очень широкое распространение. Теперь их степень влияния на взгляды людей сопоставима со степенью влияния телевидения. Диалоговый стиль общения, ограниченная ответственность авторов и огромное количество публикаций не позволяют использовать стандартные для средств массовой информации средства и методы контроля.

В данной работе рассматривается проблема выявления троллей в социальных сетях. Тролли – это пользователи социальных сетей, форумов и других дискуссионных площадок в интернете, которые нагнетают гнев, конфликт путём скрытого или явного задирания, принижения, оскорбления другого участника или участников общения. Выражается троллинг в форме агрессивного, издевательского и оскорбительного поведения [6].

Троллинг в сети наносит очень большой вред, поскольку такие пользователи могут разжигать конфликты на почве религиозной неприязни, межнациональной розни и т.п. Даже просто участие тролля в дискуссии заставляет остальных участников нервничать и тратить свое время на ответы такому пользователю. Получается, что дискуссия на какую-либо тему засоряется ненужными сообщениями.

Таким образом, задача выявления и блокирования троллей является актуальной.

## Теоретическая часть

Существуют различные методы поиска пользователей, занимающихся троллингом. Самый простой и надежный подход – это ручное модерирование дискуссий. Однако, учитывая, что в социальных сетях общаются миллионы пользователей, то ручной поиск становится слишком затратным. В этом случае, необходимо использовать методы автоматизированного поиска троллей.

В 2011 году группа исследователей из Канады разработала интересный метод выявления пользователей, которые за деньги занимались троллингом [1]. Метод основан на анализе комментариев, оставляемых троллями и их поведения в сети. Предполагается, что такие пользователи имеют схожие модели поведения, благодаря чему становится возможным их идентификация.

В таблице 1 приведен список характеристик комментариев пользователей, по которым проводился анализ.

**Таблица 1.** Список характеристик комментариев пользователей

| Характеристика | Значение |
|---|---|
| Report ID | ID нового сообщения, которое потом будет комментироваться пользователями. |
| Sequence No. | Порядковый номер комментария к сообщению. |
| Post Time | Время размещения комментария. |
| Post Location | Географическое местоположения пользователя, разместившего комментарий. |
| User ID | ID пользователя, который оставил комментарий. |
| Content | Содержимое комментария. |
| Response Indicator | Индикатор ответа показывает, является ли данный комментарий новым или ответом на другой комментарий. |

Исследователи применяли классификацию на основе семантического и несемантического анализа. При этом максимальная точность выявления составила 88.79%. Как отмечают сами исследователи, в более ранних работах, основанных только на анализе содержимого сообщений, точность не превышала 50% [2, 4].

При этом существенным недостатком является зависимость предложенного алгоритма от легко меняемых показателей, типа Post Time или Post Location. Так же ненадежным параметром является Sequence No.

На наш взгляд, наиболее точно определить, пользователь является троллем или нет, можно только анализируя текст его комментариев и частоту размещения.

Причем, учитывая тот факт, что тролли могут писать с ошибками, использовать шаблонные фразы или маскироваться под нормальных пользователей, то лучше анализировать не смысловую часть их комментариев, а эмоциональную составляющую, поскольку она гораздо труднее подделывается.

Кроме того, следует учитывать, что тролли не просто пишут комментарии, а *пытаются манипулировать другими участниками* дискуссии, что также должно проявиться в эмоциональной составляющей.

Попытаемся качественно представить себе типовые модели поведения пользователей в социальной сети при размещении комментариев.

Большая часть пользователей практически не оставляет комментариев к понравившемуся сообщению. Обычно ставят «лайк» или «класс». В лучшем случае напишут что-то вроде «супер!», «прикольно» или т.п. Т.е. мы имеем дело с одиночными комментариями, которые как правило очень короткие.

Следующая категория пользователей пишет более длинные комментарии, которые, как правило отражают их эмоциональное отношение к сообщению.

Еще одна категория участников дискуссии – это «жертвы» троллей. Как правило, они пытаются что-то доказать и при этом пишут очень развернутые комментарии с обилием эмоций.

Ну и наконец, тролли. Это люди, которые постоянно участвуют в дискуссии, пытаются спровоцировать других участников, а, следовательно, не могут отписываться короткими фразами.

Для выявления эмоционального состояния участника дискуссии мы воспользовались фактом, что структура информационного текста принципиально отличается от структуры внушающего (манипулирующего) текста и характеризуется отсутствием намеренной ритмизации его лексических и фонетических единиц [7].

На практике это означает, что некоторые звукосочетания способны не только вызывать определенные эмоции, но и могут восприниматься в качестве определенных образов [5]. Например, в сочетаниях буква «и» с указанием предмета обладает свойством «уменьшения» объекта, перед которым (или в котором) она явно доминантно присутствует. Также, звук «о» производит впечатление мягкости и расслабленности. Преобладание звуков «а» и «э», как правило, ассоциируется с эмоциональным подъемом.

Исходя из перечисленных выше предпосылок, нами были предложены для анализа поля, перечисленные в таблице 2.

**Таблица 2.** Список полей для анализа

| Поле | Комментарий | Поле | Комментарий |
|---|---|---|---|
| $M_p$ | $M_p$-количество сообщений для $p$-го пользователя | $f_{y,p}$ | Частота встречаемости символа «у» для $p$-го пользователя |
| $L_p$ | Средняя длина сообщения для $p$-го пользователя | $f_{э,p}$ | Частота встречаемости символа «э» для $p$-го пользователя |
| $f_{a,p}$ | Частота встречаемости символа «а» для $p$-го пользователя | $f_{ю,p}$ | Частота встречаемости символа «ю» для $p$-го пользователя |
| $f_{e,p}$ | Частота встречаемости символа «е» для $p$-го пользователя | $f_{я,p}$ | Частота встречаемости символа «я» для $p$-го пользователя |
| $f_{u,p}$ | Частота встречаемости символа «и» для $p$-го пользователя | $f_{!,p}$ | Частота встречаемости символа «!» для $p$-го пользователя |
| $f_{o,p}$ | Частота встречаемости символа «о» для $p$-го пользователя | $f_{?,p}$ | Частота встречаемости символа «?» для $p$-го пользователя |

Значения этих полей рассчитываются по следующим формулам:

$$f_{x,p} = \frac{\sum_i^{M_p} N_{i,x,p}}{\sum_i^{M_p} N_{i,p}} \qquad (1)$$

где $p$-порядковый номер пользователя, $i$-порядковый номер сообщения для $p$-го пользователя, $M_p$-количество сообщений для $p$-го пользователя, $x$-символ, для которого делается расчет; $N_{i,x,p}$-количество символов $x$ в $i$-м сообщении $p$-го пользователя; $N_{i,p}$-общее количество символов в $i$-м сообщении.

$$L_p = \frac{\sum_i^{M_p} N_{i,p}}{M_p} \qquad (2)$$

**Практическая часть**

Для обучения классификатора нами была подготовлена статистическая выборка из более, чем 1200 комментариев к сообщению на политическую тему. Активных участников дискуссии было выделено 145 человек. В этой выборке «вручную» были выявлены 2 тролля.

Кроме того, для тестирования была подготовлена другая выборка на религиозную тематику, состоящая из 61 комментария и 30 активных участников. Троллей в этой выборке не было.

В таблице 3 приведен пример обучающего множества.

**Таблица 3.** Пример обучающего множества

| $p$ | $M_p$ | $L_p$ | $f_{а,p}$ | $f_{е,p}$ | $f_{и,p}$ | $f_{о,p}$ | $f_{у,p}$ | $f_{э,p}$ | $f_{!,p}$ | $f_{ю,p}$ | $f_{я,p}$ | $f_{?,p}$ |
|---|---|---|---|---|---|---|---|---|---|---|---|---|
| 1 | 1 | 44 | 0,045 | 0,136 | 0,023 | 0,045 | 0,023 | 0,000 | 0,000 | 0,000 | 0,000 | 0,000 |
| 2 | 1 | 64 | 0,109 | 0,063 | 0,078 | 0,078 | 0,031 | 0,016 | 0,000 | 0,000 | 0,000 | 0,000 |
| 3 | 1 | 246 | 0,081 | 0,069 | 0,089 | 0,057 | 0,016 | 0,000 | 0,000 | 0,000 | 0,004 | 0,000 |
| 4 | 2 | 176 | 0,057 | 0,068 | 0,054 | 0,094 | 0,034 | 0,000 | 0,006 | 0,000 | 0,014 | 0,000 |
| 5 | 7 | 392 | 0,065 | 0,070 | 0,062 | 0,077 | 0,019 | 0,003 | 0,018 | 0,005 | 0,014 | 0,033 |
| 6 | 7 | 71 | 0,085 | 0,071 | 0,049 | 0,065 | 0,028 | 0,000 | 0,000 | 0,004 | 0,028 | 0,000 |
| 7 | 33 | 188 | 0,055 | 0,061 | 0,052 | 0,068 | 0,017 | 0,002 | 0,015 | 0,004 | 0,010 | 0,004 |
| 8 | 1 | 12 | 0,083 | 0,000 | 0,083 | 0,083 | 0,083 | 0,000 | 0,000 | 0,000 | 0,000 | 0,000 |
| 9 | 1 | 54 | 0,074 | 0,111 | 0,019 | 0,019 | 0,056 | 0,000 | 0,000 | 0,000 | 0,000 | 0,000 |
| 10 | 2 | 38 | 0,133 | 0,080 | 0,013 | 0,040 | 0,027 | 0,000 | 0,000 | 0,027 | 0,027 | 0,000 |
| 11 | 1 | 61 | 0,082 | 0,033 | 0,082 | 0,066 | 0,033 | 0,016 | 0,000 | 0,000 | 0,000 | 0,000 |
| 12 | 1 | 20 | 0,100 | 0,050 | 0,050 | 0,100 | 0,000 | 0,000 | 0,000 | 0,000 | 0,000 | 0,000 |
| 13 | 1 | 102 | 0,029 | 0,078 | 0,078 | 0,088 | 0,010 | 0,000 | 0,186 | 0,000 | 0,000 | 0,000 |
| 14 | 1 | 52 | 0,096 | 0,058 | 0,019 | 0,058 | 0,019 | 0,000 | 0,058 | 0,000 | 0,000 | 0,000 |

В качестве классификатора использовались самоорганизующиеся карты Кохонена [3], реализованные на базе аналитического пакета Deductor*.

Для обучения использовались следующие настройки:

1) Определение количества кластеров автоматическое;

2) Скорость обучения в начале 0,3, а в конце 0,005;

3) Радиус обучения в начале 4, а в конце 0,1;

4) Максимальное количество эпох обучения 1000.

По итогам обучения была получена карта Кохонена, представленная на рисунке 1.

---

* http://www.basegroup.ru

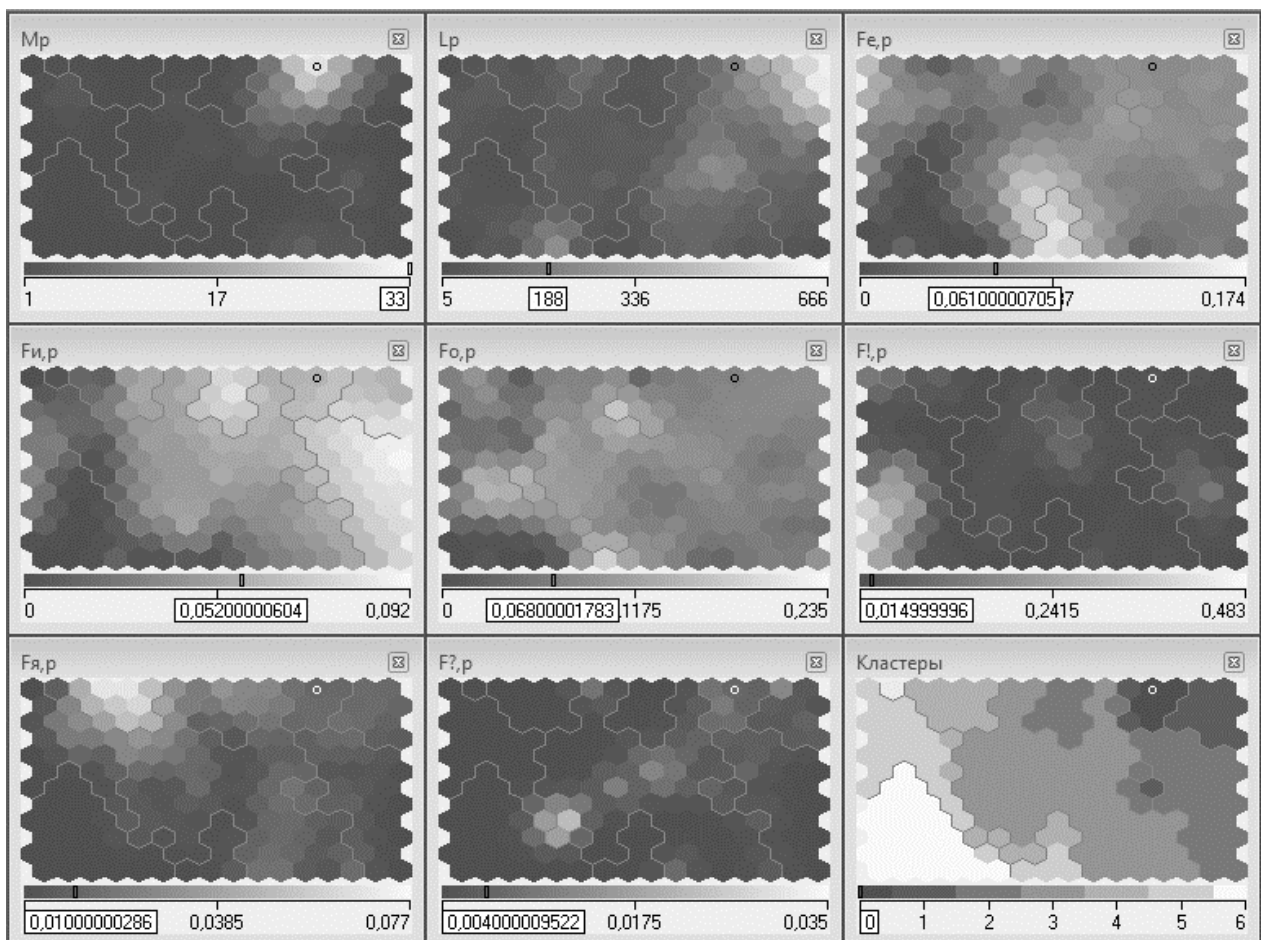

**Рис. 1.** Карта Кохонена

На данном рисунке представлены результаты построения карты Кохонена в проекциях на входные параметры. Всего было выявлено 7 кластеров.

Кластер, содержащий троллей, наиболее отчетливо виден в проекции на параметр $M_p$.

Для проверки правильности работы нейронной сети мы «вручную» выявили двух типичных троллей, причем многие участники дискуссии точно также характеризовали этих пользователей. Оба пользователя попали в кластер № 0 (на рисунке он помечен точкой). В этой же группе оказались еще два пользователя, которые после дополнительного анализа их сообщений были нами также характеризованы как тролли.

В тестовой выборке ошибочных выявлений троллей не было.

По итогам обучения была также составлена таблица значимости полей для отнесения пользователей к категории троллей.

**Таблица 4.** Таблица значимости полей

| Параметр | Значимость (в %) | Параметр | Значимость (в %) |
|---|---|---|---|
| $f_{я,p}$ | 24,7 | $f_{!,p}$ | 37,1 |
| $f_{и,p}$ | 81,0 | $f_{o,p}$ | 25,4 |
| $f_{e,p}$ | 39,2 | $f_{y,p}$ | 5,9 |
| $f_{a,p}$ | 0,0 | $f_{?,p}$ | 93,3 |
| $L_p$ | 78,3 | $f_{э,p}$ | 6,3 |
| $M_p$ | 100,0 | $f_{ю,p}$ | 7,2 |

Из данной таблицы видно, что, хотя наибольшую значимость имеет количество комментариев, но и частотные характеристики отдельных символов, определяющих эмоциональное состояние пользователя, также влияют на принятие решения об отнесении участника дискуссии к категории троллей.

На рисунке 2 представлена обобщенная схема работы предложенного нами алгоритма выявления троллей.

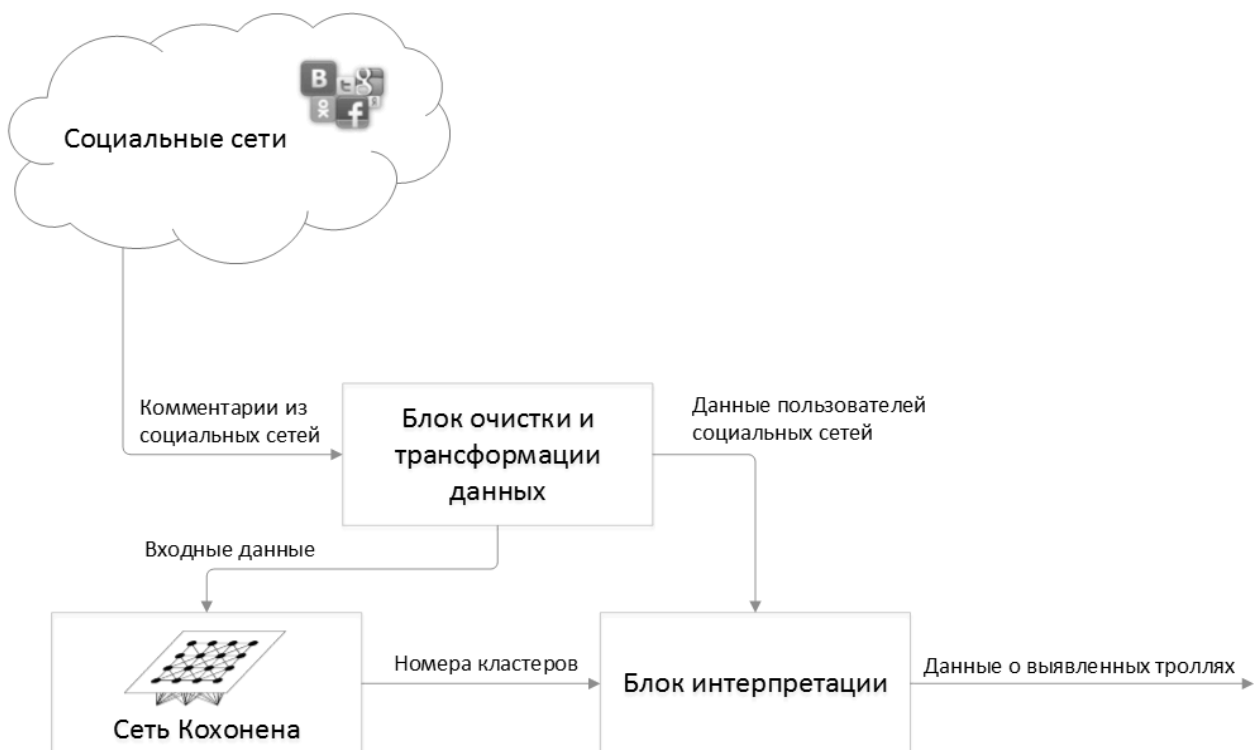

**Рис. 2.** Обобщенная схема работы алгоритма

Предложенный нами алгоритм выявления троллей состоит из следующих шагов:
1. Из выбранной дискуссии в социальной сети выбираются все комментарии;
2. Комментарии поступают в блок очистки и трансформации данных, где по формулам 1 и 2 преобразовываются к виду, представленному в таблице 3;
3. Эти данные пропускаются через сеть Кохонена;
4. На выходе сети мы получаем для каждого участника дискуссии номер кластера, к которому он относится;
5. Соотнося номера кластеров с исходными данными пользователей в блоке интерпретации, мы получаем список троллей, выявленных в данной дискуссии.

Предложенный алгоритм имеет ряд преимуществ по сравнению с описанным в начале статьи:
1. Для анализа мы не используем идентификаторы пользователей, которые легко меняются;
2. Нам не требуется учитывать географию участников дискуссии;
3. Дискуссия может быть сколько угодно растянута по времени, но это никак не скажется на качестве работы алгоритма;
4. Мы используем для анализа эмоциональную составляющую комментариев, причем выявляем эту составляющую не по словам, как у других авторов, а по частотам встречаемости тех или иных символов. Это дает устойчивость алгоритма при искажении текста или маскировке тролля.

Из недостатков следует отметить следующее:
1. Алгоритм работает эффективно, только тогда, когда дискуссия уже будет содержать несколько сотен комментариев;
2. Предложенный метод не работает, если тролли пишут единичные комментарии с разных аккаунтов или используют специальных программных ботов.

## Заключение

В данной статье был предложен метод выявления троллей в социальных сетях, который основывается на анализе эмоционального состояния пользователей сети и поведенческой активности.

Тестирование показало эффективность предложенного метода.

## Список литературы

1. C. Chen, K. Wu, V. Srinivasan, X. Zhang. Battling the Internet Water Army: Detection of Hidden Paid Posters. http://arxiv.org/pdf/1111.4297v1.pdf, 18 Nov 2011
2. D. Yin, Z. Xue, L. Hong, B. Davison, A. Kontostathis, and L. Edwards. Detection of harassment on web 2.0. Proceedings of the Content analysis in the Web, 2, 2009
3. T. Kohonen. Self-organization and associative memory. 2d ed. New-York, Springer-Verlag, 1988
4. Y. Niu, Y. min Wang, H. Chen, M. Ma, and F. Hsu. A quantitative study of forum spamming using contextbased analysis. In In Proc. Network and Distributed System Security (NDSS) Symposium, 2007
5. В.В. Киселёв. Автоматическое определение эмоций по речи. Образовательные технологии. №3, 2012, стр. 85-89
6. Р.А. Внебрачных. Троллинг как форма социальной агрессии в виртуальных сообществах. Вестник Удмуртского университета, 2012, Вып.1, стр. 48-51
7. С.В. Болтаева, Т.В. Матвеева. Лексические ритмы в тексте внушения. Русское слово в языке, тексте и культурной среде. Екатеринбург, 1997, стр. 175-185

## Сведения об авторах

**Филимонов Андрей Викторович** - к.ф.-м.н., зав. кафедрой «Прикладная информатика в экономике», Московский государственный индустриальный университет (филиал в г. Кинешме). E-mail: remueur@yandex.ru.

**Осипов Алексей Викторович** - к.ф.-м.н., доцент, кафедра «Прикладная информатика в экономике», Московский государственный индустриальный университет (филиал в г. Кинешме). E-mail: osipoffav@yandex.ru

**Климов Алексей Борисович** - к.т.н., доцент, кафедра «Прикладная информатика в экономике», Московский государственный индустриальный университет (филиал в г. Кинешме). E-mail: abxyz@rambler.ru